\documentclass[osajnl,twocolumn,showpacs,superscriptaddress,10pt]{revtex4-1}
\usepackage{graphicx}
\usepackage{colordvi}
\usepackage{amssymb}
\usepackage{amsmath}
\usepackage{bbold}
\usepackage{epsf}
\usepackage{mathrsfs} 
\usepackage{float}
\usepackage{dsfont}
\usepackage[mathscr]{euscript}
\usepackage{subfigure}

\begin{document}
\def \beq{\begin{equation}}
\def \eeq{\end{equation}}
\def \bea{\begin{eqnarray}}
\def \eea{\end{eqnarray}}
\def \bem{\begin{displaymath}}
\def \eem{\end{displaymath}}
\def \P{\Psi}
\def \Pd{|\Psi(\boldsymbol{r})|}
\def \Pds{|\Psi^{\ast}(\boldsymbol{r})|}
\def \Po{\overline{\Psi}}
\def \bs{\boldsymbol}
\def \bl{\bar{\boldsymbol{l}}}		

\title{Scattering of massless Dirac fermions in circular p-n junctions with and without magnetic field} 
\author{Neetu Agrawal (Garg)$^1$, Sankalpa  Ghosh$^2$ and Manish Sharma$^{1,3}$}
\affiliation{Centre for Applied Research in Electronics, Indian Institute of Technology Delhi, New Delhi-110016, India}
\affiliation{Department of Physics, Indian Institute of Technology Delhi, New Delhi-110016, India}
\affiliation{Atrenta India Pvt. Ltd., A-12, Sector 2, NOIDA, UP 201303, India}
\begin{abstract} 
In the absence of a magnetic field, scattered wavefunction inside a circular p-n junction in graphene exhibits an interference pattern with high intensity maximum located around the caustics. We investigate the wavefunctions in the presence of a uniform magnetic field outside the circular region to show  how the loci of the high intensity region changes by forming Landau level structure outside the circular region and a central high intensity region inside the circular p-n junction due to the strong reflection of massless Dirac fermions by the outside magnetic field. We conclude by suggesting experimental ways to detect such change of pattern due to the effect of the magnetic field. 
\end{abstract}
\pacs{72.80.Vp, 81.05.Ue, 73.63.-b, 78.20.Ci, 73.43.Qt}{}
\date{\today}
\maketitle
\section{Introduction}
Electron transport in graphene with an applied split gate voltage is akin to light propagating through a metamaterial with negative refractive index. This analogue can be used to understand the focusing of electric current by a single p-n junction in graphene and was theoretically predicted in a seminal work by Cheinov {\it et al.} \cite{cheinov}. It was shown that for the case of a symmetric n-p junctions when $(k_n = k_p)$, where $k_n(k_p)$ is the wave number in the $n(p)$ region, the n-p junction provides perfect focusing of the emitted electrons on the p-side. However, for the case of asymmetric n-p junctions when $(k_n \neq k_p)$, the sharp focus transforms into a pair of {\it caustics},  which in the language of optics corresponds to an envelope of a family of rays (wavevectors for the electrons) at which the density of rays is singular. 

The cusp-shaped patterns of light reflection on the inside of a coffee cup, the patterns of bright lines observed on the bottom of a swimming pool and the common rainbow are some prime examples of this phenomenon occurring in the nature. Since the rays in geometrical optics are analogous to classical trajectories of electrons, the concept of {\it caustics} has already been taken forward to atomic \cite{atomicrain} and nuclear scattering experiments \cite{nucrain} where an analogous to rainbow phenomena have been observed. Observation of caustics in the trajectories of cold atoms in a linear magnetic potential \cite{coldatom} have also been reported.\

\begin{figure}
\begin{center}
\centerline{\epsfxsize 7.0 cm \epsffile{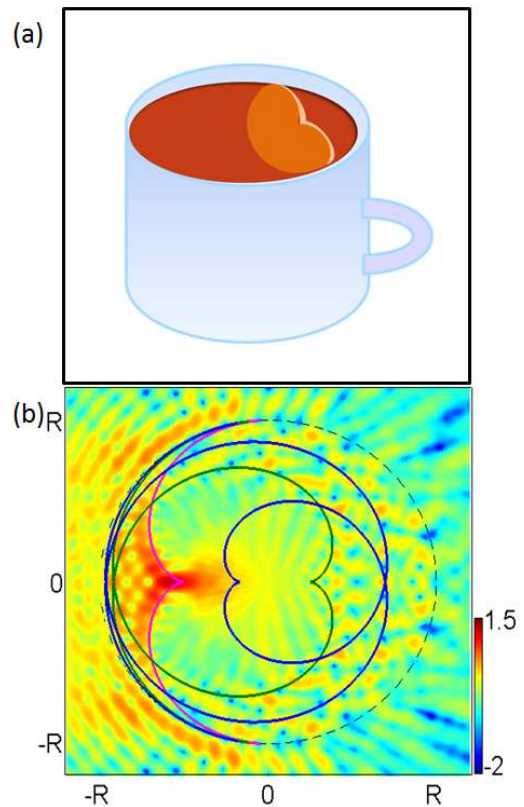}}
\end{center}
\caption{(a) Schematic coffee cup caustic effect (b) Color bar represents probability density $log_{10}|\Psi|^2$ (scale of logarithmic to base 10) distribution for a circular scalar potential barrier. x-axis and y-axis corresponds to spatial x and y in units of R, and $kR=qR=500$, $n=-1$. Line plots represents caustic curves as obtained in analogy with geometrical optics in equation \ref{paramet}}
\label{cupcaustics}
\end{figure}

Such focusing and caustic formation of charge carriers also arises in circular n-p junctions of monolayer graphene \cite{pefelvi}. The scattered wave function of an incoming plane wave of electrons due to a circular symmetric step-like potential shows an interference pattern which resembles that of `cup-caustics' \cite{berryrev} [\textit{c.f} Fig. \ref{cupcaustics}], with high intensity maximum located around the caustics that can be calculated from Snell's law with negative refractive index \cite{pefelvi}. Complex caustics patterns have also been demonstrated in the presence of Rashba spin-orbit interaction in circular gated or doping-controlled region in graphene \cite{biref}. This was shown to result in selective focusing of different spins, and the possible direct measurement of the Rashba coupling strength in scanning-probe experiments.\

Whereas the focusing and caustic formation can create region of high intensity for electrons in a circular p-n junction, one may ask what will happen to the probability distribution in presence of a suitable magnetic field. Such a query is motivated by the fact that  magnetic barriers of different geometries are known to confine the charge carriers in monolayer graphene that are massless Dirac fermions\cite{semicondscegger, peetercirc, ijmpbneetu}. Whereas the scattering problem of such massless Dirac fermions in circular potential barrier and the resulting caustic formation is studied in detail for monolayer \cite{pefelvi} and bilayer \cite{pefelvibilayer,interbandfoc} graphene, the fate of such optical analogies for the case of magnetic barriers is yet to be analyzed. This is because of the localised nature of solutions which occur in the presence of magnetic field.  

However, it may be pointed out that interesting the features like periodic spatial modulation of current injected from a point source near a p-n junction in graphene in presence of uniform magnetic field has been predicted and shown to originate from caustic bunching of skipping-snaking orbits \cite{avishkar1, avishkar2}. In the present work, we investigate the wavefunction patterns in the presence of a graphene magnetic dot. The geometry corresponds to the magnetic field which is constant except within a disk where it vanishes. One of the ways in which such an inhomogenous magnetic field can be experimentally realized is by depositing a superconducting disk on the top of graphene sheet. When a homogenous magnetic field is applied perpendicularly, the magnetic flux lines are expelled from the superconducting disk due to Meissner effect, which results in the proposed inhomogenous magnetic field profile. Other methods of creating such inhomogeous magnetic 
field for two dimensional electronic systems was recently reviewed in \cite{ijmpbneetu}. 
 For the case of a non relativistic two -dimensional electron gas (2DEG) system, such a magnetic field profile was discussed in Ref. \cite{prb59, prb64, ibrahim4} to study the bound states and analyse the optical absorption spectrum \cite{prb59} of such system. The energy spectrum for such magnetic dot in graphene has been discussed in detail in \cite{semicondscegger}. The spectrum shows a set of bound states below the lowest bulk Landau level which can be employed to design and control an artificial atom in such a graphene nanostructure. This energy spectrum is different from the energy spectrum for Schr$\ddot{o}$dinger electrons in terms of the energies of the bound states which is a consequence of the different energies of the corresponding Landau levels. It may however be noted that in ref. \cite{masir} a complementary configuration of the magnetic field was considered, where the magnetic field is nonzero only in a finite, circular disk-like region of space, and vanishes outside that region. In the same work it has been shown that in such a situation  true bound states solutions are not possible and only quasi bound states can be defined. 
Thus this configuration provides very different spectrum as compared to the case under consideration in this paper. 
Given this significant change in the energy spectrum in the presence of the proposed magnetic field profile we investigate that what will happen to the caustic formation of the Dirac electron wave functions of monolayer graphene in circular p-n junctions.

The paper is organised as follows. We begin with understanding the scattering of incident plane wave of ballistic electrons on a circular potential step and obtain the interference pattern which resembles `cup caustics'  that can be described in terms of geometrical optics. In the next section, for the case of a magnetic dot, we consider the electrons to be incident (in all directions) from inside the dot which is the field free region, and analyze the interference patterns. The results and discussion as well as the description in terms of classical treatment is then presented in the section thereafter.

\section{Theory}
\subsection{Circular n-p junction}
We consider charge carriers in graphene in presence of a circular gate potential
\beq V({\bf r}) = V_0\Theta(R-r) \eeq
In the absence of intervalley scattering, the low-energy dynamics for graphene charge carriers is described by the following Dirac-like Hamiltonian in plane polar co-ordinates 
\beq \textbf{H} = -i\hbar v_F \left[ \begin{matrix} 0 & e^{-i\phi}\left(\frac{\partial}{\partial r} - \frac{i}{r}\frac{\partial}{\partial \phi}\right) \\e^{i\phi}\left(\frac{\partial}{\partial r} + \frac{i}{r}\frac{\partial}{\partial \phi}\right) & 0 \end{matrix} \right] + V({\bf r})\mathds{1} \label{hamv}\eeq
\noindent
For rotationally symmetric scalar potential form like the one which is chosen here, pseudoangular momentum operator  $J_z = -i\hbar\partial_\phi+\hbar\sigma_z/2$ commutes with the Hamiltonian. 
\beq [J_z, {\bf H}] = 0 \eeq
For this reason the wavefunction solutions can be constructed in the following form \cite{bookkatnelson}:
\bea \Psi_1(r,\phi) &=& e^{il\phi}\Psi_1(r) \\
\Psi_2(r,\phi) &=& e^{i(l+1)\phi}\Psi_2(r) \label{trial}\eea
where $l \in \mathbb{Z} = \{...-1, 0, 1...\} $ is pseudo-angular momentum which is a conserved quantity here. On substituting Eq. \ref{trial} to solve for ${\bf H} \Psi = E\Psi$ in Eq. \ref{hamv}, we obtain the following set of coupled equations
\bea \left[\frac{\partial }{\partial r}- \frac{l}{r}\right]\Psi_1(r) = i\frac{E-V}{\hbar v_F}\Psi_2(r) \\
\left[\frac{\partial }{\partial r}+ \frac{l+1}{r}\right]\Psi_2(r) = i\frac{E-V}{\hbar v_F}\Psi_1(r) \eea
\noindent
On decoupling the above set of equations we obtain
\beq (kr)^2\frac{\partial^2\Psi_1}{\partial (kr)^2} + (kr)\frac{\partial\Psi_1}{\partial (kr)} + [(kr)^2-l^2]\Psi_1 = 0\eeq
\noindent
l being an integer, the general solution for the above equation are given as
\beq \Psi_1(r) = c_1J_l(kr) + c_2Y_l(kr) \nonumber \eeq
\noindent
But since $Y_l(kr)$ diverges as $r \rightarrow 0$ so instead one can use Hankel's functions 
\beq H_l^{(1,2)}(kr) = J_l(kr) \pm iY_l(kr) \nonumber \eeq
with the asymptotics at $kr>>1$
\beq H_l^{(1,2)}(kr) \approx \sqrt{\frac{2}{\pi k r}}\exp{\left[\pm i\left( kr+\frac{l\pi}{2}-\frac{\pi}{4}\right )\right]} \eeq
Thus, the function $H_l^{(1)}$ describes the scattering wave and $H_l^{(2)}$ describes the wave falling at the centre. By expanding the incident plane wave (taken parallel to $\phi=0$ axis) in terms of eigenfunctions in polar coordinates, the net wavefunction at $r>R$ can be given as \cite{bookkatnelson}
\beq \Psi_{r>R} =  \sum_{l = -\infty}^{\infty} \left( \begin{matrix} i^l[J_l(kr)+b_lH_{l}^{(1)}(kr)]e^{il\phi} \\ s i^{l+1} [J_{l+1}(kr)+b_lH_{l+1}^{(1)}(kr)]e^{i(l+1)\phi}  \end{matrix} \right),\label{wfout}\eeq
\noindent
where $s = sgn(E)$, $k=\frac{E}{\hbar v_F}$ and the terms proportional to Bessel(Hankel) functions describe incident(scattering) waves. 
\noindent
At $r<R$, k should be replaced by
\beq q= \frac{|E-V_0|}{\hbar v_F}\eeq
and only Bessel functions $J_l(qr)$ are allowed (otherwise, the solution will not
be normalizable, due to divergence of $Y_l (z)$ at $z \rightarrow 0$). Then
\beq \Psi_{r<R} =  \sum_{l = -\infty}^{\infty} a_l\left( \begin{matrix} i^lJ_l(qr)e^{il\phi} \\ s^{'} i^{l+1} J_{l+1}(qr)e^{i(l+1)\phi}  \end{matrix} \right) \label{wfin} \eeq
\noindent
Here $s^{'} = sgn(E-V_0)$ and the complex factors $b_l$ and $a_l$ in Eq. (\ref{wfout}) and (\ref{wfin}) are scattering coefficients. Upon matching the wavefunctions at the boundary $r = R$, 
\bea b_l &=& \frac{J_l(qR)J_{l+1}(kR)-(s^{'}/s)J_{l+1}(qR)J_{l}(kR)}{(s^{'}/s)H_l^{(1)}(kR)J_{l+1}(qR)-J_{l}(qR)H^{(1)}_{l+1}(kR)}  \label{tlexp} \\
a_l &=& \frac{H_{l+1}^{(1)}(kR)J_{l}(kR) - J_{l+1}(kR)H_{l}^{(1)}(kR)}
{H_{l+1}^{(1)}(kR)J_{l}(qR) -(s^{'}/s)J_{l+1}(qR)H^{(1)}_{l}(kR)} \label{alexp} \eea

\noindent
Using the above expressions (\ref{wfout}), (\ref{wfin}), (\ref{tlexp}), and (\ref{alexp}) we evaluate the probability density can be plotted as shown in Fig. \ref{cupcaustics}(c). For the scattering of an incident plane wave on a circular p-n junction, the region of high probability density can be seen to make a certain pattern which is similar to that of a `cup caustics' [{\it c.f} Fig. \ref{cupcaustics}]. \\

\noindent
In the following section we evaluate analytically the equations governing these caustics pattern which is in accordance with an analogy with geometrical optics.

\subsection{Recapitulating caustics formation in n-p junction}
In accordance with an analogy with geometrical optics, a ray incident at an impact parameter $b =  R\sin\alpha$ is refracted at an angle $\beta  = sin^{-1}\left(\frac{\sin\alpha}{|n|}\right)$.
Incident rays with different impact parameters ($-R < b < R$) form a family of rays inside the circle. If a family of rays (for different impact parameters) is traced,
they form a curved envelope. This envelope is the caustic. This, can be found from the condition that for a ray incident at an impact parameter $b = R\sin\alpha$, any point $(x_c, y_c)$ lying on the refracted ray, at a ray length $w$, the mapping  
$(w,\alpha) \rightarrow (x_c, y_c)$ is singular. Mathematically, this is given by the condition that 
\beq \det \mathscr{J} = 0 \label{detjeq0} \eeq

\begin{figure}[t]
\centerline{\epsfxsize 7.0 cm \epsffile{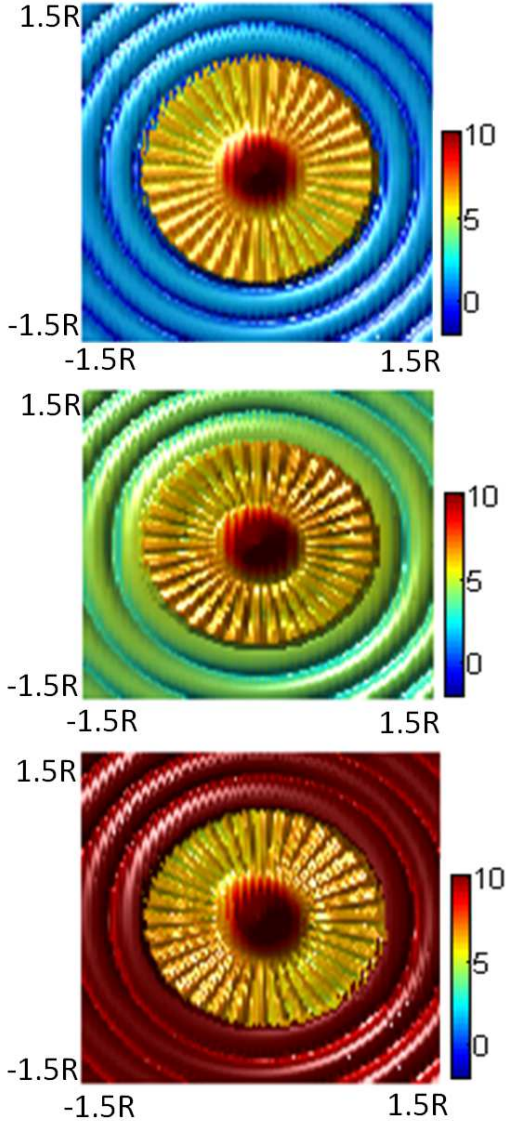}}
\caption{Probability density $log_{10}|\Psi|^2$ (scale of logarithmic to base 10) distribution for a magnetic dot geometry. $E=50meV$. x-axis and y-axis corresponds to spatial x and y in units of R, and (a) $B =  0.05 T$, (b) $B = 0.5 T$, (c) $B=1 T$ respectively.}
\label{bcausticsrayfig}
\end{figure}

For a ray undergoing single reflection from the circular n-p junction, 
\bea x_c &=& -R\cos\alpha+w\cos(\alpha+\beta) \nonumber \\
y_c &=& R\sin\alpha-w\sin(\alpha+\beta)  \eea
\noindent
so that the condition (\ref{detjeq0}) determines 
\beq w = \frac{R\cos\beta}{1+\beta '} \nonumber \eeq
The envelop as determined with this mapping is plotted (magenta curve) in Fig.\ref{cupcaustics}(b). For a ray undergoing $p-1$ internal reflections from the n-p junction, the envelop is determined as \cite{pefelvi},

\bea \frac{r_c}{R} &=& (-1)^{p-1}\left[\left(\begin{matrix} -\cos\zeta \\ \sin\zeta \end{matrix}\right)\right. \nonumber \\
&+& \left.\cos\beta\frac{1+2(p-1)^{\beta^{'}}}{1+(2p-1)\beta^{'}} \left(\begin{matrix} \cos(\zeta+\beta) \\ -\sin(\zeta +\beta)\end{matrix}\right) \right] \label{paramet}\eea
where 
\bea \zeta &=& \alpha+2(p-1)\beta  \nonumber \\
\beta^{'} &=& \frac{\cos\alpha}{\sqrt{n^2-\sin^2\alpha}} \eea
This is a parametric curve for caustics and is plotted in Fig. \ref{cupcaustics}(b) for $p=1$ (magneta curve), 2 (green curve), 3 (blue curve). In both cases, we see that the caustics pattern formed from the wavefunction probability distribution obtained using full quantum mechanical calculation inside the junction agrees very well with that obtained from Snell's law with negative refractive index. In the next section we investigate such probability density patterns for the case of a magnetic dot.

\subsection{Magnetic dot}

In this section we investigate the effect of magnetic field ${\bf B} = B_0\Theta(r-R)$ on the scattering pattern in such n-p junction. We begin with calculating the wavefunctions in the presence of magnetic dot profile as mentioned above. For such radially symmetric magnetic field, adopting the symmetric gauge, the radial and azimuthal components of the vector potential are 
\beq  A_r(r) = 0; ~~ A_\phi(r) = \frac{r^2-R^2}{2r}\Theta(r-R) \eeq

\noindent
Then, upon solving for $ {\bf H}\Psi = E\Psi $ for the Hamiltonian 
{ \small
\bea && {\bf H}=\nonumber \\
&&\left[ \begin{matrix} 0 & e^{-i\phi}\left(-i\frac{\partial}{\partial r} - \frac{1}{r}\frac{\partial}{\partial \phi}+A_r-iA_\phi\right) \\e^{i\phi}\left(-i\frac{\partial}{\partial r} + \frac{1}{r}\frac{\partial}{\partial \phi}+A_r+iA_\phi \right) & 0 \end{matrix} \right] \nonumber\eea
}
\noindent
the solutions inside the dot are obtained as Bessel's functions while outside the dot the solutions can be written in terms of confluent hypergeometric solutions. This is shown in detail below. As explained in the previous section, the eigen states of the above written Hamiltonian  are classified by
\beq \Psi_{j}(r,\phi) = \left(\begin{matrix}\Phi_j(r)e^{i(j-1/2)\phi}\\ \chi_j(r)e^{i(j+1/2)\phi}\end{matrix}\right) \eeq
\noindent
On substituting for $\Psi_{j}(r,\phi)$ in  $ {\bf H}\Psi = E\Psi $, we obtain two coupled equations as:
\bea \left[\frac{d}{dr} + \left(\frac{j+1/2}{r} + A_\phi \right)\right]\chi_j &=& \frac{E}{\hbar v_F}\Phi_j \nonumber \\
\left[\frac{d}{dr} - \left(\frac{j-1/2}{r} + A_\phi \right)\right]\Phi_j &=& \frac{E}{\hbar v_F}\chi_j \label{set1equse}\eea
\noindent
Upon solving the above set of equation for the $r<R$, the decoupled equation is obtained as below \cite{semicondscegger}
\beq \frac{d^2\Phi_j}{dr^2}+\frac{1}{r}\frac{d\Phi}{dr}+\left[k^2-\frac{(j-1/2)^2}{r^2}\right]\Phi_j = 0 \eeq
\noindent
of which the solutions can be written in terms of Bessel's functions $J_l(kr)$ as discussed for scalar barrier in the previous section as well.
\beq \Psi_j(r,\phi) \propto \left(\begin{matrix} J_{j-1/2}(kr)e^{i(j-1/2)\phi} \\ i J_{j+1/2}(kr)e^{i(j+1/2)\phi} \end{matrix}\right) \label{inside}\eeq
\noindent
Similarly outside the dot, using $\xi = r^2/2$, equations (\ref{set1equse}) reduces to the following pair of coupled linear differential equations,
\bea \left[\frac{d}{dr} + \frac{1}{2}+\left(\frac{\check{j}+1/2}{2\xi} + A_\phi \right)\right]\chi_j &=& \frac{k}{\sqrt{\xi}}\Phi_j \\
\left[\frac{d}{dr} - \frac{1}{2}-\left(\frac{\check{j}-1/2}{2\xi} + A_\phi \right)\right]\Phi_j &=& \frac{k}{\sqrt{\xi}}\chi_j \eea
\noindent
where $\check{j} = j-R^2/2$ accounts for the missing magnetic flux. Here we can recognize the solutions satisfied by confluent hypergeometric functions and are obtained as \cite{semicondscegger}

\begin{widetext}

For $\check{j}>0$ \
\beq \Psi_{\check{j}}(r,\phi) \propto  e^{-\xi/2}\xi^{\frac{1}{2}|\check{j}+\frac{1}{2}|}\left(\begin{matrix} \frac{k}{\sqrt{\xi}}M(a,|\check{j}+\frac{1}{2}|;\xi)e^{i(\check{j}-1/2)\phi} \\ i \frac{k}{\sqrt{\xi}}M(a,|\check{j}+\frac{1}{2}|+1; \xi)e^{i(\check{j}+1/2)\phi} \end{matrix}\right) \eeq

For $\check{j}<0$ \
\beq \Psi_{\check{j}}(r,\phi)  \propto e^{-\xi/2}\xi^{\frac{1}{2}|j+\frac{1}{2}|}\left(\begin{matrix} k\sqrt{\xi}M(a'+1,|\check{j}+\frac{1}{2}|+2;\xi)e^{i(\check{j}-1/2)\phi} \\ i \frac{k}{\sqrt{\xi}}M(a',|\check{j}+\frac{1}{2}|+1; \xi)e^{i(\check{j}+1/2)\phi} \end{matrix}\right) \eeq
\end{widetext}
\noindent
where $M(a,\check{j}; \xi)$ corresponds to the confluent hypergeometric functions and
\bea a &=& |\check{j}+\frac{1}{2}|-k^2 \\
&a'& = -k^2 \eea

\noindent
We consider the charge carriers to be incident (in all directions) from inside the dot (i.e $B=0$ region) to outside. However as shown in Eq. (\ref{inside}), the solutions inside the dot are localised Bessel's functions. Thus, in order to carry out the analysis we decompose the Bessel's functions in the following form of asymptotic $(z>>1)$,


\bea J_\nu(z) &\sim& \left(\frac{2}{\pi z}\right)^{1/2}\left[\cos\omega \sum_{m=0}^{\infty}\frac{(-1)^m(\nu,2m)}{(2m)^{2m}}\right. \nonumber\\
&-& \left. \sin\omega\frac{(-1)^m(\nu,2m+1)}{(2z)^{2m+1}}\right] \eea
where 
\beq \omega = z-\frac{1}{2}\nu-\frac{1}{4}\pi \nonumber \eeq
and \beq (\nu, m) = (-1)^m\frac{(\frac{1}{2}-\nu)_m(\frac{1}{2}+\nu)_m}{m!} = \frac{\Gamma (\nu+m+\frac{1}{2})}{m! \Gamma (\nu-m+\frac{1}{2}) }\nonumber \eeq
\noindent
This decomposes the wavefunction inside the dot in terms of radially incoming and outgoing waves, so that close to the boundary we can assume the charge carriers to be incident from inside the dot towards outside.\\
\begin{figure}[t]
\begin{center}
\centerline{\epsfxsize 5.0 cm \epsffile{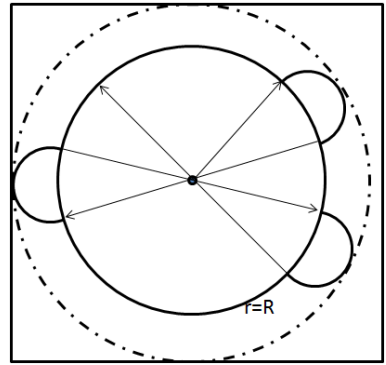}}
\end{center}
\caption{Figure depicting classical trajectory for the electrons incident from inside the dot towards outside, the wavevector follows a straight line in the field free region inside the dot beyond which it encounters a continuously varying vector potential, the radial momentum vanishes at classical turning points as calculated from Eq. (\ref{turnout}), $m_l = 0$.}
\label{semiclassicalfig5}
\end{figure}

\noindent
Next, by matching the solutions at the boundary of the dot $r=R$, we obtain the complete normalised wavefunctions both inside and outside the dot. 

\begin{figure}[t]
\begin{center}
\centerline{\epsfxsize 7.0 cm \epsffile{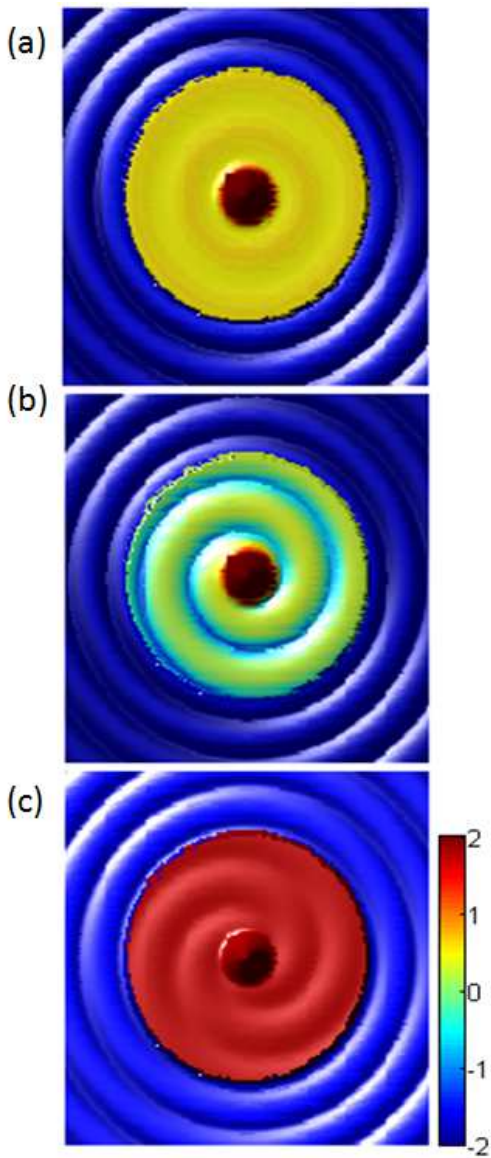}}
\end{center}
\caption{Partial wave analysis: Probability density  $log_{10}|\Psi|^2$ (scale of logarithmic to base 10) distribution for (a)s-wave (b) p-wave, (c) d-wave, for a magnetic dot geometry. $E=50meV$, x-axis and y-axis corresponds to spatial x and y varying from $-1.5 R$ to $1.5 R$, and $B = 0.5 T$.}
\label{pwfig}
\end{figure}

\section{Results and Discussion}
\noindent
By matching the solutions at the boundary of the dot $r=R$, we obtain the complete normalised wavefunctions both inside and outside the dot. The probability density distribution $|\Psi(r,\phi)|^2$ (on logarithmic base 10 scale) is plotted in Fig. \ref{bcausticsrayfig}. Clearly the high probability density curves corresponds to circular regions only. This can be understood in terms of classical treatment as well in the following manner. \\

\noindent
Since the vector potential inside the dot is zero so classically, for the electrons incident from inside the dot towards outside, the  wavevector follow a straight line till the boundary where it encounters a continuously varying vector potential and hence refractive index. The classical trajectory bends continuously till its radial momentum vanishes [c.f Fig. \ref{semiclassicalfig5}]. The classical energy-momentum relation of a massless particle moving in the fields A and V , with energy E and radial momentum p, is

\beq p^2 = \frac{(E-V)^2}{(\hbar v_F)^2} -\left(\frac{m_l}{r}+eA_\phi\right)^2 \eeq
\noindent
where the term $\frac{m_l}{r}$ is due to the angular momentum, $m_l = 0, \pm 1, \pm 2 ... $. The classical motion is restricted to the region where $p^2 > 0$, and $p^2 =0$ defines the classical turning points \cite{tunable, classical}. For our magnetic dot geometry, we obtain the classical turning points as follows.
Inside the dot 
\beq r = \pm \frac{m_l}{E/\hbar v_F} \label{insideturn}\eeq
and outside the dot there will be two turning points obtained using the following quadratic equation
\beq \frac{m_l}{r}+\frac{eB}{\hbar c}\frac{r^2-R^2}{2r} = \pm \frac{E}{\hbar v_F} \label{turnout}\eeq
\noindent
Expression (\ref{insideturn}) implies that for s- wave corresponding to $m_l = 0$, the classical trajectories always pass through the centre of the dot. This will not be the case for higher order $m_l>0$ contributions [c.f. Fig. \ref{pwfig}].\\

\begin{figure}[t]
\begin{center}
\centerline{\epsfxsize 6.9 cm \epsffile{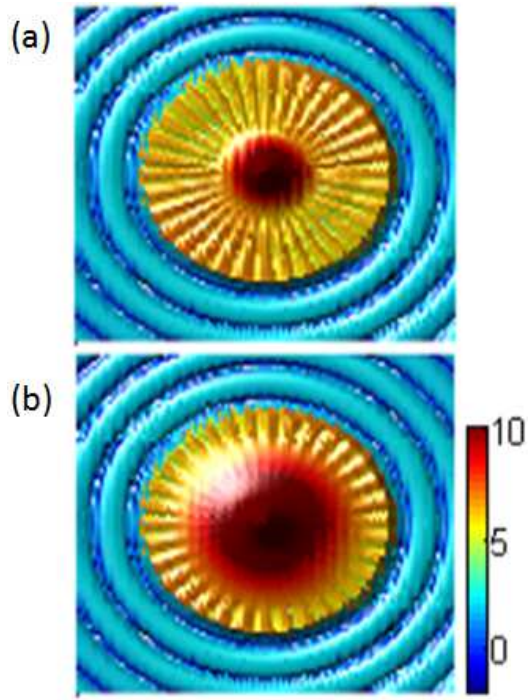}}
\end{center}
\caption{Probability density $log_{10}|\Psi|^2$ (scale of logarithmic to base 10) distribution for a magnetic dot geometry with a commensurate scalar potential. $E=50meV$, x-axis and y-axis  corresponds to spatial x and y varying from $-1.5 R$ to $1.5 R$, and $B = 0.001 T$, (a) $V= 0$, (b) $V=75meV$}
\label{voltageb0causticsrayfig}
\end{figure}

\noindent
As against the scalar dot in graphene where we see a cup-caustics like features, here the probability density plot does not show any such feature. This is also because of the underlying symmetry of the problem where the electrons are incident from inside the dot towards outside in all directions. Unlike the scalar case the incident flux cannot be chosen in one direction from outside the barrier because the vector potential of the magnetic field extends upto infinity, also the solutions outside the dot are localised unlike the plane wave solutions for the case of scalar barrier. As the magnetic field is turned on the $2m_l+1$ degenerate states are pulled to the closest Landau level (subband) of quantum number $n$. Consequently, the Landau levels can be visualized by the circles as can be clearly seen from Fig. \ref{bcausticsrayfig}, and the probability density at the  Landau levels increases with increasing magnetic field. However the probability density distribution inside the dot does not change to a large extent with changing magnetic field outside the dot. This is because the refractive index inside the dot remain same even if we change the magnetic field outside. \\

\noindent
This situation however changes if a scalar potential $V$ is present inside the dot. The classical turning inside the dot in this case depends on the scalar potential as well. 
\beq r = \pm \frac{m_l}{(E-V)/\hbar v_F} \eeq 
\noindent
The effective refractive index inside the dot changes due to the presence of scalar potential $V$, this appears in the probability density distribution plots as well as the size of the high probability density regions changes with changing scalar potential. This is clearly depicted in Fig. \ref{voltageb0causticsrayfig} as shown above.\\

\noindent
In conclusion we have analysed the wavefunction probability density distribution for a magnetic dot using a full quantum mechanical calculation and explained using classical treatment as well. The solutions inside the dot are localised Bessel's functions in nature, using a particular form for asymptotic we decompose these localised wavefunctions in terms of radially incoming and outgoing waves, thereby completely determining the wavefunctions inside and outside the dot. The caustic features which appear in the presence of a scalar dot in graphene are no more present in the presence of a magnetic dot. Our results as presented may be useful for understanding of scanning-gate microscopy on graphene quantum nanostructures.  It was  pointed out in earlier studies  on non relativistic 2DEG\cite{prb59}, that such magnetic field profile can be realized following the work A. K. Geim {\it et. al} \cite{geimnatexp, geimaplexp}. Here the superconducting disks with radius ranging from $0.25-1.2$ micrometer were studied in a magnetic field in the range $50-100$ Gauss. The superconducting disks were placed on top of a $GaAs/Al_xGa_{1-x}As$ heterostructure and the separation between the 2DEG and the superconducting disk was about $100$ nm. Other methods of creating such nonuniform magnetic field profile using hard ferromagnetic material was also discussed in a recent review\cite{ijmpbneetu}. These suggest that a graphene magnetic dot with well calculated parameters can be experimentally realized. Such magnetic dot coupled with source and drain leads via two constrictions can be used for scanning gate experiments. For illustration, experimentally Ref. \cite{imaging} has shown the imaging of resonant states of the quantum dot and the constrictions in real space. Also, Ref. \cite{scanprl} has employed the scanning tunnelling spectroscopy to study the real-space local density of states of a two dimensional electron system in a magnetic field. We hope that our results provide some insight and may augment the future work in this direction.

\section{Acknowledgment}
This work is supported by grant SR/S2/CMP-0024/2009 of DST, Govt. of India. One of the authors (NA) acknowledges financial support from CSIR, Govt. of India.

\end{document}